\newif\ifconfver
  \confverfalse      
 \confvertrue        

\ifconfver
	\documentclass{article}
	\usepackage{spconf}
	\title{One-Bit Massive MIMO Precoding via a Minimum Symbol-Error Probability Design}
 	\name{Mingjie Shao$^{\dagger}$, Qiang Li$^{\ddag}$ and Wing-Kin Ma$^{\dagger}$	
\thanks{This work was supported in part by the National Natural Science Foundation of China under Grant 61531009, and in part by the Fundamental Research Funds for the Central Universities under Grants ZYGX2016J011.}
}
	\address{
	$^{\dagger}$ Department of Elec. Eng., The Chinese University of Hong Kong, Hong Kong SAR, China	 \\
	$^{\ddag}$ School of Info. \& Comm. Eng., University of Electronic Science and Technology of China, China
	}
\else
	\documentclass[11pt]{article}
	\usepackage{spconf}
\title{One-Bit Massive MIMO Precoding via a Minimum Symbol-Error Probability Design}
	\author{Mingjie Shao$^{\dagger}$, Qiang Li$^{\ddag}$ and Wing-Kin Ma$^{\dagger}$	
\thanks{This work was supported in part by the National Natural Science Foundation of China under Grant 61531009, and in part by the Fundamental Research Funds for the Central Universities under Grants ZYGX2016J011.}}
\address{
	$^{\dagger}$ Department of Elec. Eng., The Chinese University of Hong Kong, Hong Kong SAR, China	 \\
	$^{\ddag}$ School of Info. \& Comm. Eng., University of Electronic Science and Technology of China, China
	}
\fi
\usepackage{amsmath,graphicx}

\usepackage{calc,amsfonts,amssymb,bm,url,color,theorem,cite}
\usepackage{psfrag,subfigure,float}
\usepackage{algorithm}
\usepackage{algorithmic}
\usepackage{mathtools,lipsum,cuted,multicol}
\usepackage{filecontents}

\definecolor{orange}{RGB}{255,107,0}

\newtheorem{Lemma}{Lemma}

\theorembodyfont{\rmfamily}

\newcommand\bs{\ensuremath{{\bm s}}}
\newcommand\bv{\ensuremath{{\bm v}}}
\newcommand\bw{\ensuremath{{\bm w}}}

\newcommand\bh{\ensuremath{{\bm h}}}

\newcommand\bu{\ensuremath{{\bm u}}}

\newcommand{\setD}{\mathcal{D}}
\newcommand{\setX}{\mathcal{X}}

\newcommand{\Rbb}{\mathbb{R}}
\newcommand{\Cbb}{\mathbb{C}}

\newcommand{\Exp}{\mathbb{E}}

\newcommand\bx{\ensuremath{{\bm x}}}
\newcommand\by{\ensuremath{{\bm y}}}
\newcommand\bY{\ensuremath{{\bm Y}}}
\newcommand\bH{\ensuremath{{\bm H}}}

\newcommand\bn{\ensuremath{{\bm n}}}
\newcommand\bS{\ensuremath{{\bm S}}}
\newcommand\bN{\ensuremath{{\bm N}}}
\newcommand\bX{\ensuremath{{\bm X}}}
\newcommand\bz{\ensuremath{{\bm z}}}

\newcommand\bbx{\ensuremath{\bar{\bm x}}}
\newcommand\bbs{\ensuremath{\bar{\bm s}}}

\newcommand{\Rfrak}{\mathfrak{R}}
\newcommand{\Ifrak}{\mathfrak{I}}

\makeatletter
\def\bstctlcite{\@ifnextchar[{\@bstctlcite}{\@bstctlcite[@auxout]}}
\def\@bstctlcite[#1]#2{\@bsphack
  \@for\@citeb:=#2\do{%
    \edef\@citeb{\expandafter\@firstofone\@citeb}%
    \if@filesw\immediate\write\csname #1\endcsname{\string\citation{\@citeb}}\fi}%
  \@esphack}
\makeatother

\begin{document}
%
\bibliographystyle{IEEEtran}

\ifconfver
	\ninept
\fi

\bstctlcite{IEEEexample:BSTcontrol}
\maketitle
\begin{abstract}
Massive multiple-input multiple-output (MIMO) has the potential to substantially improve the spectral efficiency, robustness and coverage of mobile networks. However, such potential is limited by hardware cost and power consumption associated with a large number of RF chains. Recently, one-bit quantization is proposed to address this issue by replacing high-resolution digital-to-analog converters (DACs) with one-bit DACs, thereby  simplifying the RF chains. Despite low system cost, advanced signal processing techniques are needed to compensate for quantization distortions caused by low-resolution DACs. In this paper, a symbol-error-rate (SER)-based one-bit precoding scheme is proposed to minimize the detection error probability of all users under one-bit constraints. The problem is recast as a continuous optimization problem with a biconvex objective. By applying the block coordinate descent (BCD) method and the FISTA method, we develop an efficient iterative algorithm to obtain a one-bit precoding solution. Simulation results demonstrate its superiority over state-of-the-art algorithms in terms of bit error rate performance in high-order modulation cases.
\end{abstract}

\section{Introduction}
\label{sec:intro}

Multiple-input multiple-output (MIMO) has been well known to be a powerful, and almost indispensable, technique to achieve high spectral efficiency in modern communication systems \cite{Goldsmith2003}. Massive MIMO, where the base station (BS) is equipped with hundreds or even thousands of antennas, exhibits more desirable properties such as robustness to channel fading and high energy efficiency \cite{Rusek2013,Lu2014}, and is essential for novel systems such as millimeter-wave systems \cite{Pi2011}.
However, hardware cost and power consumption associated with increased number of RF chains poses a serious limit on how massive we can practically deploy massive MIMO. A major part of power consumption and hardware cost comes from high-resolution digital-to-analog converters (DACs) \cite{Murmann2017}. The use of one-bit DACs serves as a potential approach for circumventing this problem \cite{Risi2014}.
However, the coarse quantizations of one-bit DACs bring about great challenges in precoding design. Direct implementation of  linear precoding  by quantization suffers from considerable performance loss. This paper focuses on the precoding design in one-bit massive multiuser MIMO downlink to mitigate the effect of multiuser interference and quantization distortions.

\par One-bit quantization using low-precision analog-to-digital converters (ADCs) was first analyzed for massive MIMO uplink \cite{Choi2015,Choi2016,Studer2016}, while research for one-bit MIMO precoding is quite limited.  Methods based on linear precoding with quantization effects taken into account were proposed \cite{Swindlehurst2016,Mezghani2009}. Although linear quantized precoding methods deliver reasonable performance in high SNR regions when the number of transmit antennas is large and the symbol constellation density is low \cite{Swindlehurst2016}, they perform poorly when denser symbol constellations are used.
Nonlinear precoding methods emerged recently. Nonlinear precoding methods in general operate in a per-symbol time manner,  making use of both channel state information (CSI) and symbol information.
In \cite{Swindlehurst2017}, it was shown that simple combinations of perturbation and linear precoding can effectively improve the symbol error rate (SER). In \cite{Jacobsson2017}, one-bit precoding was formulated as a symbol minimum mean-square error (MMSE) problem, which was empirically shown to give better bit error rate (BER) performance than linear quantized procoding. The MMSE method can also be extended to high-order modulation schemes, but it suffers from error floor effects in dense constellations such as $16$-QAM and $64$-QAM \cite{Jacobsson2016}.
Based on an approximate formulation of \cite{Jacobsson2017}, two low-complexity designs were proposed in \cite{Castaneda2017} to reduce the complexity at the cost of performance loss.
The idea of constructive interference for one-bit precoding was also considered in \cite{Jedda2017} for PSK signalings.
Although these works intend to reduce the SER as seen in their designs, the underlying relation between their formulations and SER remains unclear.

\par In this paper, we focus on the one-bit massive multiuser MIMO downlink precoding design with general QAM signaling. Starting from symbol error rate (SER) analysis, we propose a new formulation to  minimize the SER under the one-bit constraints. The optimization problem turns out to be a minimax problem. Noticing that dense constellations are vulnerable to amplitude mismatches caused by quantization distortions, the binary constraints must be carefully treated in the precoding design.
Instead of using simple relaxation, we propose to reformulate the problem as a continuous optimization problem with a biconvex objective via a variational reformulation of the binary constraints.
The resulting optimization problem is non-convex and non-smooth, which is challenging to tackle. By noticing that the variables can be divided into two blocks, and the optimization problem with respect to each block is convex, we apply block coordinate descent (BCD) to update the two block variables. In particular, the optimization with respect to one of two blocks has a closed-form solution, while  that of the other block can be updated with custom-derived fast algorithm. Simulation results reveal that our formulation leads to significant BER performance gains compared to the state of the art.\\[-1.5em]

\section{System Model and Problem Formulation}
\label{sec:format}

Consider a single-cell massive multiuser MIMO block-fading downlink scenario,
where the BS exploits the CSI and symbol information to precode multiple data streams, one for each single-antenna user, simultaneously. The received signal model over one transmission block is
\begin{equation} \label{model}
    y_{i,t}=\bh_i^T \bx_t + n_{i,t} ,~~i=1,\ldots, K,~~ t=1,\ldots,T.
\end{equation}
Here, $\bx_t \in \setX$ is the multi-antenna transmitted signal at symbol time $t$; $\setX$ is the feasible set of $\bx_t$, and under the one-bit DAC constraints it takes the form
 \[
 \setX=\left\{\bx\in \Cbb^{N}~|~x_i=\sqrt{\frac{P}{2N}}(\pm 1 \pm j)\right\}
 \]
 where $N$ is the number of transmit antennas, ranging from hundreds to thousands in a typical massive MIMO system;
 $P$ is the total transmission power of the BS; $T$ is the length of one transmission block;
 $y_{i,t}$ is the received signal of user $i$ at symbol time $t$;
 $K$ is the number of users; $\bh_{i} \in \Cbb^{N}$ is the downlink channel from the BS to user $i$, which remains unchanged within the transmission block;
 $n_{i,t}\sim \mathcal{CN}(0, \sigma_n^2)$ is additive white Gaussian noise. The signal model \eqref{model} can be concisely written as
\[\bY=\bH \bX +\bN,
\]
 where $\bY=[\by_1,\by_2,\ldots, \by_T]$, $\bH=[\bh_{1},\bh_{2},\ldots, \bh_{K}]^T$, $\bX=[\bx_1,\bx_2, \ldots, \bx_T]$ and $\bN=[\bn_1,\bn_2,\ldots, \bn_T]$.  We aim to design $\{\bx_t\}_{t=1}^{T}\in \setX$ such that the impacts of quantization error and multiuser interference are minimized.
\par To put into context, let $\bs_t=[s_{1,t}, s_{2,t},\ldots , s_{K,t}]^T$ be the symbol vector sent at symbol time $t$. Also, let $\bS=[\bs_1, \ldots, \bs_T]$. The symbols $s_{i,t}$'s are drawn from a QAM constellation $\mathcal{S}$, which is defined as
\[
    \mathcal{S}=\{ s_{R}+j s_{I}~|~s_{R},s_{I}\in \{ \pm 1,\pm 3,\ldots, \pm (2L-1) \}\},
\]
where $L$ is the order of the QAM constellation. Our task is to shape desired symbols at the user sides. To be explicit, we seek to achieve
\[
    \bh_i^T \bx_t \approx d \cdot s_{i,t},
\]
where $d\geq 0$ is a signal gain factor.
 The detection at user $i$ is
\[
    \hat{s}_{i,t}=\text{dec}(y_{i,t}/ d),
\]
where the $\text{dec}(\cdot)$ is the decision function of $\mathcal{S}$. Note that $d$ is assumed to be known at the users, which can be achieved via training.
The symbol error probability of user $i$ at symbol time $t$ is denoted as
\begin{equation*}
  \begin{split}
    \text{SEP}_{i,t}=\text{Pr} (\hat{s}_{i,t} \neq s_{i,t}| s_{i,t}).
  \end{split}
\end{equation*}
It is easy to verify that
\begin{equation*}
\begin{split}
\text{SEP}_{i,t} \leq \text{SEP}_{i,t}^{R} + \text{SEP}_{i,t}^{I}\leq 2\max
\{ \text{SEP}_{i,t}^{R}, \text{SEP}_{i,t}^{I} \},
\end{split}
\end{equation*}
where $\text{SEP}_{i,t}^{R}=\text{Pr}(\Rfrak\{\hat{s}_{i,t}\}\neq \Rfrak\{s_{i,t}\}|s_{i,t})$
    denotes the probability that an error occurs in the direction of the in-phase component, while $\text{SEP}_{i,t}^{I}=\text{Pr}(\Ifrak\{\hat{s}_{i,t}\}\neq \Ifrak\{s_{i,t}\}| s_{i,t})$ stands for that in the direction of the quadrature component. It can be proved that
\begin{equation*}
  \begin{split}
    \text{SEP}_{i,t}^{R}&\leq \underbrace{ 2 Q\left(  \frac{d-|\Rfrak\{\bh_i^T \bx_t\}-d\Rfrak\{s_{i,t}\} | }{\sigma_n/ \sqrt{2}} \right)}_{M^{R}_{i,t}},\\
    \text{SEP}_{i,t}^{I} &\leq \underbrace{ 2 Q\left( \frac{d-|\Ifrak\{\bh_i^T \bx_t\}-d\Ifrak\{s_{i,t}\} | }{\sigma_n/ \sqrt{2}} \right)}_{M^{I}_{i,t}},
  \end{split}
\end{equation*}
where $Q(x)=\int_{x}^{\infty} \frac{1}{\sqrt{2\pi}} e^{-z^2/ 2} dz$.
\par Our one-bit precoding design is to attempt to minimize users' SERs in the worst-case sense. Specifically, using the SEP upper bounds above, we consider the following design formulation
\begin{equation}\label{minmax}
\begin{split}
\min_{\bx_t,d}&~~ \max_{i,t} ~2\max\left\{M^{R}_{i,t},{M^{I}_{i,t}}\right\}\\
\text{s.t.}& ~~\bx_t\in \setX,d \geq 0,~~ i=1,\ldots,K,~~t=1,\ldots, T.
\end{split}
\end{equation}
Problem \eqref{minmax} can be notationally simplified to
\begin{equation}\label{realMinmax}
\begin{split}
    \min_{\bbx_t,d }&~~ \max_{i,t} ~~4 Q\left(\frac{d-|\bar{\bh}_i^T \bar{\bx}_t-d\bar{s}_{i,t} | }{\sigma_n/\sqrt{2}} \right)\\
    \text{s.t.}& ~~\bbx_t\in \setX_{\Rfrak},d \geq 0,~~ i=1,\ldots,2K,~~t=1,\ldots, T,
\end{split}
\end{equation}
where $\setX_{\Rfrak}=\{\bx\in \Rbb^{2N}~|~x_i=\pm \sqrt{\frac{P}{2N}} \}$ is the real-valued equivalent feasible set;
\begin{gather*}
\bar{\bH}=[\bar{\bh}_1, \ldots, \bar{\bh}_{2K}]^T=\begin{bmatrix}
               \Rfrak\{\bH\} & -\Ifrak\{\bH\} \\
               \Ifrak\{\bH\} & \Rfrak\{\bH\}
             \end{bmatrix};\\
             \bar{\bX}=[\bbx_1,\ldots, \bbx_T]=\begin{bmatrix}\Rfrak(\bX)\\\Ifrak(\bX) \end{bmatrix};
             ~~  \bar{\bS}=[\bbs_1,\ldots,\bbs_T]=\begin{bmatrix}
                \Rfrak\{\bS\} \\
                \Ifrak\{\bS\}
              \end{bmatrix}.
\end{gather*}
Since $Q(\cdot)$ is a monotonically decreasing function, problem \eqref{realMinmax} is equivalent to
\begin{equation} \label{Form1}
  \begin{split}
    \min_{\bbx_t,d }&~~\max_{t} ||\bar{\bH} \bar{\bx}_t-d\bar{\bs}_{t}||_{\infty}-d,\\
    \text{s.t.}&~~ \bar{\bx}_t \in \setX_{\Rfrak},~~ d\geq 0,~~t=1,\ldots, T.
  \end{split}
\end{equation}
For notational convenience, we further rewrite problem \eqref{Form1} as
\begin{equation}\label{BKvec}
  \begin{split}
    \min_{\bar{\bx},d }&~~||\hat{\bH} \bar{\bx}-d\bar{\bs}||_{\infty}-d\\
    \text{s.t.}&~~ \bar{\bx} \in \bar{\setX},~~ d\geq 0.
  \end{split}{}
\end{equation}
where $\bar{\bx}={\rm vec}(\bar{\bX})$, $\bar{\setX}=\{\bx\in \Rbb^{2NT}~|~x_i=\pm \sqrt{\frac{P}{2N}} \}$, $\bar{\bs}={\rm vec}(\bar{\bS})$ and  $\hat{\bH}=\mathbf{I}_K\otimes \bar{\bH}$.
\par Our challenge is to solve the one-bit precoding problem in \eqref{BKvec}, a non-convex problem with binary constraints.
One way to deal with the binary constraints is to apply box relaxation, where the binary constraints are relaxed as intervals and the box relaxed solution is quantized to yield an approximate binary solution.
Here, we use a variational reformulation of binary constraints of which the box relaxation can be seen as a special case:
\begin{Lemma} \label{lem:1}
Consider the following optimization problem
\begin{equation}\label{binary}
  \min_{\bx \in \Rbb^{n}} ~ f(\bx) ~~\text{s.t.}~~ \bx \in \{ -1,+1 \}^n,
\end{equation}
where $f$ is an $L$-Lipschitz continuous convex function on $-\mathbf{1}\leq \bx \leq \mathbf{1}$. Given $\lambda >2L$, problem \eqref{binary} is equivalent to the following optimization problem
\begin{equation}\label{MPEC}
  \begin{split}
    \min_{\bx\in \Rbb^{n},\bv\in \Rbb^{n}}&~~ f(\bx)+\lambda (n-\bx^T \bv)~~\\
    \text{s.t.~}& -\mathbf{1}\leq \bx \leq \mathbf{1},~||\bv||_2^2\leq n.
  \end{split}
\end{equation}
Moreover, at the optimal solution $(\bx^{\star},\bv^{\star})$ to problem \eqref{MPEC}, it holds that $\bx^{\star}=\bv^{\star} \in \{ -1,+1 \}^n$.
\end{Lemma}
Lemma \ref{lem:1} is a direct consequence of Lemma $1$ and Theorem $1$ in \cite{Yuan2016}, and we omit the proof here. By Lemma \ref{lem:1}, we may reformulate problem \eqref{BKvec} as
\begin{equation}\label{ICA}
  \begin{split}
    \min_{\bbx, d, \bv}&~~F_{\lambda}(\bar{\bx}, \bv, d)\triangleq||\hat{\bH} \bar{\bx}-d\bar{\bs}||_{\infty}-d +\lambda ( PT-\bar{\bx}^T\bv)\\
    \text{s.t.}&~~ -\sqrt{\frac{P}{2N}}\mathbf{1}\leq \bar{\bx} \leq \sqrt{\frac{P}{2N}}\mathbf{1},~~||\bv||_2^2\leq  PT,~ d \geq 0.
  \end{split}
\end{equation}
\par By now, we have transformed a binary constrained optimization problem into a continuous biconvex optimization problem with the aid of $\bv$ and $\lambda$.
Note that it always holds that $\bbx^T \bv-PT \leq 0$ for any feasible $(\bbx, \bv)$, and the optimal solution satisfies $\bbx^{\star~T} \bv^{\star}-PT=0$. Moreover, if $\lambda=0$, problem \eqref{ICA} is a convex relaxation of problem \eqref{BKvec}. By increasing $\lambda$, we gradually enforce $\bbx^T \bv-PT=0$, resulting in a binary solution.
In the next section, we will build an efficient algorithm tailored for problem \eqref{ICA}.

\section{One-bit Precoding Algorithm}
\label{sec:pagestyle}

In the previous section, a biconvex formulation is proposed for the one-bit precoding problem. Now we establish efficient algorithms for problem \eqref{ICA}.
 Notice that problem \eqref{ICA} is convex in $(\bar{\bx},d)$ given $\bv$, and  convex in $\bv$ given $(\bbx,d)$. We apply the block descent method to solve problem \eqref{ICA}. The variables are divided into two blocks, specifically, $(\bbx,d)$ and $\bv$. The parameter $\lambda$ is iteratively increased to force the satisfaction of $\bbx^T \bv=1$.  In the beginning, $\lambda$ is set very small in order to get a reasonable starting point. This is summarized in Algorithm \ref{BCD}.
\begin{algorithm}[H]
\caption{: BCD Method for Solving Problem \eqref{ICA}}
\begin{algorithmic}[1]
\STATE Initialize  $\bv^{(0)}=\mathbf{0}$, $\lambda>0$, $\delta>1$, $m=0$, $M$
\REPEAT
    \STATE Update $[\bar{\bx}^{(m+1)},d^{(m+1)}]=\arg\min\limits_{\bar{\bx},d}~ F_{\lambda}(\bar{\bx}, \bv^{(m)},d)$;
    \STATE Update $\bv^{(m+1)}=\arg\min\limits_{\bv}~ F_{\lambda}(\bar{\bx}^{(m+1)}, \bv,d^{(m+1)})$;
    \STATE Update $\lambda= \lambda\times \delta$ every $M$ iterations;
    \STATE  $m=m+1$;
\UNTIL some convergence criterion is satisfied.
\end{algorithmic}\label{BCD}
\end{algorithm}


In each step, Algorithm 1 deals with a convex subproblem. The $\bv$ subproblem in Step $4$ is reduced to
\[
\bv^{(m+1)}=\arg \min_{||\bv||_2^2\leq PT}~~- (\bbx^{(m+1)})^T \bv .
\]
If $\bbx^{(m+1)}=\mathbf{0}$, then any feasible $\bv$ is an optimal solution; if $\bbx^{(m+1)}\neq \mathbf{0}$, then the optimal $\bv$ is given by
\[
    \bv^{(m+1)}=\sqrt{PT}~\bar{\bx}^{(m+1)}/||\bar{\bx}^{(m+1)}||_2.
\]
As a result, the update for $\bv$ has a closed-form solution.\\
For the $(\bbx,d)$ subproblem, Step 3 is updated by solving a convex problem. Let us write down the optimization problem in Step $3$ as follows
\begin{equation}\label{step3}
  \begin{split}
    \min_{\bar{\bx},d}&~~||\hat{\bH} \bar{\bx}-d\bar{\bs}||_{\infty}-d +\lambda ( PT-\bar{\bx}^T \bv^{(m)})\\
    \text{s.t.}&~~ -\sqrt{\frac{P}{2N}}\mathbf{1}\leq \bar{\bx} \leq \sqrt{\frac{P}{2N}}\mathbf{1}, ~~ d\geq 0.
  \end{split}
\end{equation}
Problem \eqref{step3} is a large-scale non-smooth problem. Prevailing methods to handle such a problem include the alternating direction method of multipliers (ADMM) method, proximal gradient method and its accelerated versions. To take advantage of nice structure of the constraints in \eqref{step3}, we decide to develop an accelerated proximal gradient method --- a FISTA \cite{Beck2009} type method.
Basically, consider the following problem
\[
	\min_{\bx \in \setD}~f(\bx)
\]
where $f$ is a convex function with Lipschitz continuous gradient, and $\setD$ is a convex set. At iteration $l$, the update of $\bx$ in FISTA takes the form
\begin{equation*}
	\begin{split}
	\bu^{l}=&~\bx^{l}+\frac{t_l-1}{t_{l+1}}(\bx^{l}-\bx^{l-1}),\\
		\bx^{l+1}=&~ \Pi_{\setD} (\bu^{l}-\gamma_l \nabla f(\bu^{l})),
	\end{split}
\end{equation*}
where $t_{l+1}=\frac{1+\sqrt{1+4t_l^2}}{2}$; $\Pi_{\setD}$ is the projection operator onto set $\setD$; $\gamma_l$ is a step size, which can be determined either by computing a Lipschitz constant of $\nabla f$, or by a backtracking line search \cite{Beck2009}.

FISTA is computationally efficient if $\Pi_{\setD}$ is easy to compute. Moreover, FISTA guarantees a faster convergence rate $\mathcal{O}(1/l^2)$ than the classical gradient projection method $\mathcal{O}(1/l)$.
\begin{algorithm}[htb!]
\caption{: FISTA for Problem \eqref{LSE}}
\begin{algorithmic}[1]
\STATE Initialize  $\bz^{0}=\bz^{-1}=(\bar{\bx}^{0},d^{0})$, $\sigma$ in $f$, $t_0=0$
\REPEAT
    \STATE Find $\gamma_l$ via backtracking line search
    \STATE Compute
\begin{equation*}
\begin{split}
t_{l+1}&=\frac{1+\sqrt{1+4 t_{l}^2}}{2};\\
\bw^{l}&=\bz^{l}+\frac{t_{l}-1}{t_{l+1}}(\bz^{l}-\bz^{l-1});\\
	\bz^{l+1}&= \Pi_{\setD} (\bw^{l}-\gamma_l \nabla f(\bw^{l}));
\end{split}
\end{equation*}
where $\nabla f(\bbx,d) = [\frac{\partial f}{\partial \bar{\bx}};\frac{\partial f}{\partial d}]$ and
\begin{equation*}
 \begin{split}
\frac{\partial f}{\partial \bar{\bx}}&=\frac{\sum_{i=1}^{2KT}\exp\left[\frac{(\hat{\bh}_i^T \bar{\bx}-d \bar{s}_{i})^2}{\sigma} \right] (\hat{\bh}_i^T \bar{\bx}-d\bar{s}_{i}) \hat{\bh}_i}{\sqrt{\sigma \log W}\times W}-\lambda \bv^{(m)},\\
\frac{\partial f}{\partial d}&=\frac{-\sum_{i=1}^{2KT}\exp\left[\frac{(\hat{\bh}_i^T \bar{\bx}-d\bar{s}_{i})^2}{\sigma} \right] (\hat{\bh}_i^T \bar{\bx}-d\bar{s}_{i}) \bar{s}_{i}}{\sqrt{\sigma \log W}\times W}-1,
\end{split}
\end{equation*}
with $W=\sum_{i=1}^{2KT}\exp\left[\frac{(\hat{\bh}_i^T \bar{\bx}-d\bar{\bs}_{i})^2}{\sigma}\right]$;
\STATE $l=l+1$;
\UNTIL {some stopping criterion is satisfied.}
\end{algorithmic}\label{FISTA}
\end{algorithm}
\par  Let us apply FISTA to problem \eqref{step3}.  As seen in problem \eqref{step3},  we have $\setD=\{(\bar{\bx},d)| - \sqrt{\frac{P}{2N}}\mathbf{1}\leq \bar{\bx} \leq \sqrt{\frac{P}{2N}}\mathbf{1},~~ d\geq 0\}$. The projection $\Pi_{\setD}$ can be easily evaluated by thresholding.
The challenge for applying FISTA type method to handle problem \eqref{step3} lies in the non-smoothness of the objective function. We start by using the LogSumExp (LSE) function to smoothen the infinity norm, resulting in the following problem
\begin{align}\label{LSE}
    \min_{\bar{\bx}, d}&~f (\bar{\bx}, d)\! \!\triangleq \!\!\sqrt{\!\sigma \log \!\sum_{i=1}^{2KT}\!\exp\!\! \left[\!\frac{(\hat{\bh}_i^T \bar{\bx}-d\bar{\bs}_{i})^2}{\sigma}\!\right]} \!\!- \!d \!+\!\lambda ( PT\!-\bar{\bx}^T \bv^{(m)}\!)\notag\\
    \text{s.t.}&~~ -\sqrt{\frac{P}{2N}}\mathbf{1}\leq \bar{\bx} \leq \sqrt{\frac{P}{2N}}\mathbf{1},~~ d\geq 0,
\end{align}
where $\sigma$ is the smoothing parameter; note that $\text{LSE}(\bx)\rightarrow \max\{ \bx \}$ for $\sigma \rightarrow 0$. Thus, using a small $\sigma$ can approximate problem \eqref{step3} well.
Since the Lipschitz constant of $\nabla f(\bbx,d)$ is not easy to compute in our case, we apply FISTA with backtracking line search. The pseudo-code of our developed FISTA type algorithm is shown in Algorithm \ref{FISTA}.
\section{Simulation Results}
\label{sec:typestyle}

In this section, we show the simulation results of our proposed algorithm in Algorithm \ref{BCD}. To benchmark our algorithm, we have compared our algorithm with zero-forcing (ZF) with/without quantization and the state-of-the-art squared $l_{\infty}$-norm relaxation algorithm (SQUID) \cite{Jacobsson2017} method.

 The channel considered here is a block Rayleigh fading channel. The BS has $N=128$ transmit antennas. There are $K=16$ single-antenna users. We will use $16$-QAM and $64$-QAM in the simulation. The length of one transmission block is $T=10$. BER is used as the performance metric in all simulations.
  Algorithm \ref{BCD} stops when $\lambda$ is larger than twice of the Lipschitz constant of the objective in problem \eqref{BKvec}. The reported BERs are averaged BERs over 10,000 independent channel realizations.

\begin{figure}[htb!]
\centering
\includegraphics[width=\linewidth]{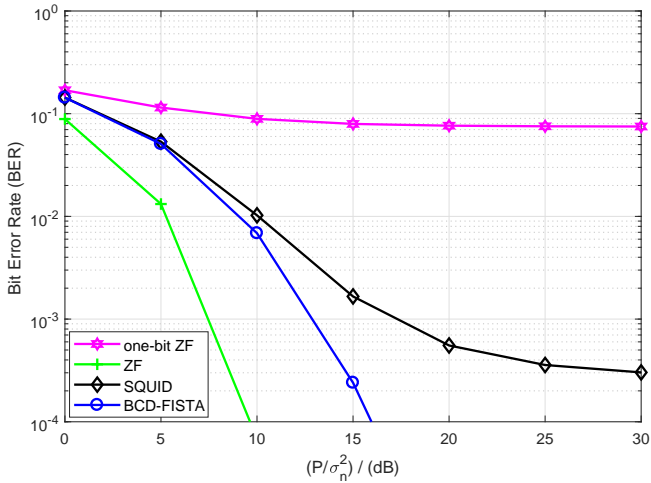}
\caption{Average BER performance versus $P/\sigma_n^2$; $16$-QAM.}\label{16sim}
\end{figure}

\begin{figure}[htb!]
\centering
\includegraphics[width=\linewidth]{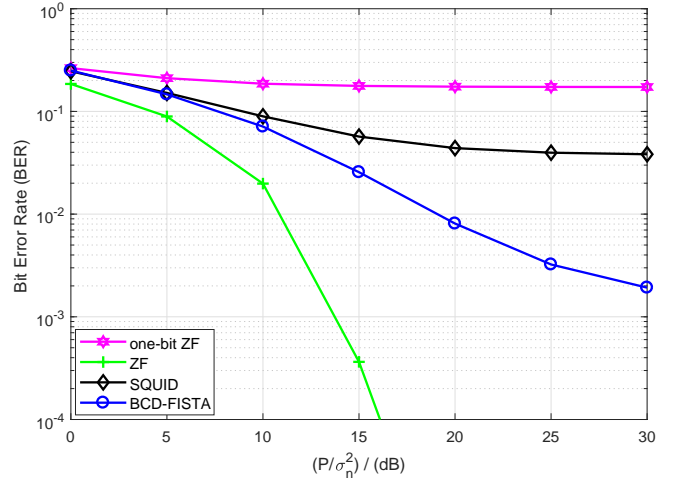}
\caption{Average BER performance versus $P/\sigma_n^2$; $64$-QAM.}\label{64QAM}
\end{figure}

\par Fig. \ref{16sim} shows the BER performance when $16$-QAM is used for transmission. In the legend, ``ZF''  stands for the zero-forcing precoding under the transmission power constraint $\Exp[||\bx_t||_2^2]=P$ with high-resolution DACs, while ``one-bit ZF'' stands for the one-bit quantization version of ``ZF''; ``SQUID'' represents the nonlinear precoding algorithm in \cite{Jacobsson2017}; ``BCD-FISTA'' is our proposed one-bit precoding algorithm, with smoothing parameter $\sigma=0.01$. It is seen that both ``SQUID'' and ``BCD-FISTA'' outperform ``one-bit ZF''. Also, ``BCD-FISTA'' achieves a much better BER performance than ``SQUID''. The performance gap between ``ZF'' with high resolution DACs and ``BCD-FISTA'' is about $5$dB  when BER$=10^{-3}$. Fig. \ref{64QAM} illustrates the simulation result  under $64$-QAM modulation.  We see that ``BCD-FISTA'' significantly outperforms ``SQUID'' for $64$-QAM.
Also, we notice that ``SQUID''  suffers from error floor when the SNR is high.
\begin{figure}[htb!]
\centering
\includegraphics[width=\linewidth]{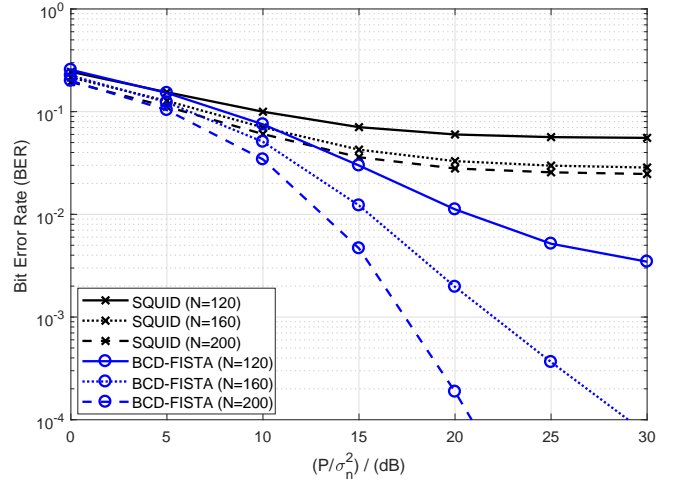}
\caption{Average BER performance versus $P/\sigma_n^2$ with different number of antennas.}\label{antennas}
\end{figure}

Fig. \ref{antennas} shows the impact of the number of transmit antennas. There are $K=16$ users and the number of transmit antennas ranges from $120$ to $200$. The $64$-QAM signaling is used for transmission. We compare the BER performance for ``BCD-FISTA'' and ``SQUID'' when the number of transmit antennas increases. We see that ``BCD-FISTA'' benefits a lot from increasing the number of transmit antennas, while ``SQUID'' is much less sensitive to. Thus, our algorithm enjoys a favorable scaling property.
 In addition, we should mention that our proposed algorithm is much faster than ``SQUID''. For example, it is $2.6\times$ and $3.5\times$ faster than ``SQUID'' under the configuration of $N=160$ and $N=200$, respectively.

\section{Conclusion}
\label{conc}
In this paper, we have proposed an SER-based precoding formulation for one-bit massive MIMO with general QAM signalings. Also, we have developed a custom-designed algorithm to handle the resulting biconvex problem. Simulation results showed that the proposed method significantly outperformed state of the art with much lower computational complexity.

\vfill\pagebreak
\newpage

\bibliographystyle{IEEEtran}
\bibliography{refs}
\nocite{*}

\end{document}